\documentclass[aps,pre,twocolumn,showpacs]{revtex4}
\usepackage{graphicx,color,epsfig}
\usepackage{amssymb,amsmath,amsfonts}
\begin{document}

\title{Amplitude death in coupled robust-chaos oscillators}
\author{M. J. Palazzi} 
\author{M. G. Cosenza}
\affiliation{Grupo de Investigaci\'on en Caos y Sistemas Complejos, Centro de F\'isica Fundamental, \\
Universidad de Los Andes, M\'erida, M\'erida 5251, Venezuela}

\begin{abstract}  
We investigate the synchronization behavior of a system of globally coupled, continuous-time oscillators possessing robust chaos.
The local dynamics corresponds to the Shimizu-Morioka model where the occurrence of robust chaos in a region of its 
parameter space has been recently discovered.  
We show that the global coupling can drive the oscillators to synchronization into a fixed point created by the coupling, 
resulting in amplitude death in the system. 
The existence of robust chaos allows to introduce heterogeneity in the local parameters,
while guaranteeing the functioning of all the oscillators in a chaotic mode. 
In this case,  the system reaches
a state of oscillation death, with coexisting clusters of oscillators in different steady states. 
The phenomena of amplitude death or oscillation death in coupled robust-chaos flows could be employed as mechanisms 
for stabilization and control in systems that require reliable operation under chaos. 
\end{abstract}

\maketitle

A number of practical uses  of the phenomenon of chaos have been proposed in recent
years, as for example, in secure communications \cite{Hayes,Argy}, in cryptography \cite{Rosa}, control of nonlinear systems \cite{Bocca,Uchida}, 
in mixing in chemical processes \cite{Ottino}, 
in avoiding electromagnetic interferences \cite{Deane}, in stabilizing plasma fusion \cite{Chandre}, etc. 
In such applications it is of importance to obtain reliable operation of chaotic systems.

It is known that most chaotic attractors of smooth systems are embedded with a
dense set of periodic windows for any range of parameter values. Therefore in practical
systems functioning in a chaotic regime, a slight fluctuation of a parameter may drive the system out of chaos. 
On the other hand, it has been found that some dynamical systems can exhibit robust chaos \cite{Baner,Potapov,Priel}. A
chaotic attractor is said to be robust if, for its parameter values, there exist a neighborhood in the parameter space with absence of
periodic windows and the chaotic attractor is unique \cite{Baner}. 
Robustness is an important property in applications that require reliable
operation in a chaotic regime, in the sense that the chaotic behavior cannot be destroyed by arbitrarily small perturbations of the system
parameters. For instance, networks of coupled maps with robust chaos have efficiently been used in communications schemes \cite{Garcia}.

Most examples of systems displaying robust chaos correspond to discrete maps \cite{Kawabe,Javier,Ali,Alvarez,Sprott}. Recently, 
the existence of robust chaos in a continuous time three-dimensional system 
have been reported by Gallas \cite{Gallas}. This system corresponds to a version of 
the Shimizu-Morioka model of a homogeneously broadened single-mode laser \cite{Shimizu}, and is given by
\begin{eqnarray}
\label{Shi}
 \frac{dx}{dt} &=& y ,\nonumber \\
  \frac{dy}{dt} &=& x(1-z)-by ,\\
   \frac{dz}{dt} &=& -a(z-x^2) .\nonumber 
\end{eqnarray}
Gallas showed that robust chaos occurs in an extensive region of the space of parameters $(a,b)$ \cite{Gallas}. 

In the search for practical applications, it is of interest to study diverse processes occurring in dynamical systems 
in the context of robust chaos. In particular, synchronization constitutes a fundamental phenomenon in interacting nonlinear 
systems \cite{Pikovsky}. An important emergent phenomenon in this context, called amplitude death, has recently attracted much attention 
\cite{Eli,Mat,Reddy,Rama1,Atay,Konishi,Prasad,Rama2,Koseska}.
Amplitude death occurs when, as a consequence of interactions, the state variables of a system get synchronized into a fixed point 
in its phase space, and it can be of relevance as a control mechanism in oscillatory or chaotic dynamics of 
various physical and biological systems \cite{Rama2,Koseska}.

In this paper we investigate the phenomenon of amplitude death in a system possessing robust chaos dynamics.
Specifically, we consider the following system of globally coupled oscillators, whose individual dynamics is given by Eq.~(\ref{Shi}),
\begin{eqnarray}
  \label{Homo}
 \frac{dx_i}{dt} &=& y_i ,\nonumber \\
  \frac{dy_i}{dt} &=& x_i(1-z_i)-by_i,\\  
   \frac{dz_i}{dt} &=& -a(z_i-x_i^2)  +\epsilon \bar{z},\nonumber 
\end{eqnarray}
where $x_i(t), y_i(t), z_i(t)$ describe the state variables of oscillator $i=1,2,\ldots,N$, at time $t$; the parameter 
$\epsilon$ measures  strength of the global coupling between oscillators; and where
we define the mean values
\begin{eqnarray}
\bar{z}(t)&=&  \frac{1}{N} \sum_{j=1}^N z_j(t) , \\
\bar{y}(t)   &=&   \frac{1}{N} \sum_{j=1}^N y_j(t) ,\\
\bar{x}(t) &=&  \frac{1}{N} \sum_{j=1}^N x_j(t) .
\end{eqnarray}
The additive  coupling term employed in Eq.~(\ref{Homo}) is one of the simplest functional forms of global coupling 
that leads to amplitude death for the given local dynamics. 

\begin{figure}[h]
\begin{center}
\includegraphics[width=0.7\linewidth,angle=0]{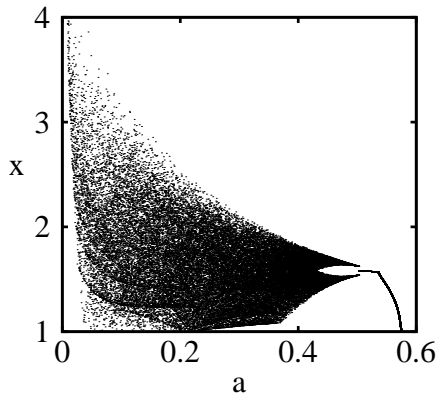}
\end{center}
\caption{Bifurcation diagram of the values $x>1$ on the plane $z=1$ intersecting the attractor of Eq.~(\ref{Shi}), as a function of $a$. 
The corresponding values of $b$ are given by Eq.~(\ref{recta}). For each value of $a$, $400$ consecutive values of $x$ have been plotted, 
after discarding a transient time of $3 \times 10^3$ integration steps.}
\label{F1}      
\end{figure}

In order to set the local parameters in Eq.~(\ref{Homo}) in the robust chaos regime, we shall consider values of $a$ and $b$ 
given by the straight line
\begin{equation}
\label{recta}
 b=1.57 a+0.2 ,
\end{equation}
which intersects the wide robust chaos region on the space of parameters $(a,b)$ of the system Eq.~(\ref{Shi}). 
To show the existence of robust chaos in the local dynamics for parameter values $a$ and $b$ satisfying Eq.~(\ref{recta}), we consider a projection 
of the attractor of system Eq.~(\ref{Shi}) on the plane $z=1$, and take the values of the variable $x$ such that $x>1$. 
Then, we construct 
a bifurcation diagram of those values of $x$ as a function of the parameter $a$, as shown in Fig.~(\ref{F1}),
with the corresponding values of $b$ given by the Eq.~(\ref{recta}).
The absence of periodic windows on a large interval of $a$ is evident.

A synchronized, stationary state  corresponds to $x_i(t)=x^*$,  $y_i(t)=y^*$,  $z_i(t)=z^*$, $\forall i$, where the values $x^*,y^*,z^*$ 
are constants. The coupled system Eq.~(\ref{Homo}) possesses three of such states, corresponding to the values 
\begin{eqnarray}
 x^*=\pm \sqrt{1-\frac{\epsilon}{a}}, \;  y^*=0, \;  z^*=1, & \; \mbox{for} \; \epsilon < a  , \\
 x^*=0,  \; y^*=0,  \; z^*=k, & \; \mbox{for} \; \epsilon=a ,
\end{eqnarray}
where $k$ is a constant that depends on initial conditions. 
For values $\epsilon>a$, the values of the state variables $x_i(t), y_i(t), z_i(t)$, $\forall i$, increase and eventually diverge to infinity.  

The occurrence of stable synchronization in the system Eq.~(\ref{Homo}) can be numerically characterized by the asymptotic time-average
$\langle \sigma \rangle$ (after discarding a transient time) of the instantaneous standard deviations
of the distribution of state variables, defined as
\begin{equation}
\sigma (t)  =   \left[  \frac{1}{N} \sum_{i=1}^N   (x_i- \bar x)^2 + (y_i- \bar y)^2 + (z_i- \bar z)^2 \right]^{1/2} .
\end{equation}
Then, a synchronization state corresponds to a value $\langle \sigma \rangle=0$. 

We have numerically integrated the system Eq.~(\ref{Homo}) with size $N=10^4$ and fixed parameter values $a=0.375$ and $b=0.826$, 
both on the robust chaos region of the local dynamics and satisfying Eq.~(\ref{recta}), for different values of the coupling parameter $\epsilon$.
A fourth-order Runge-Kutta scheme with fixed integration step $h=0.01$ was used. 

\begin{figure}[h]
\begin{center}
\includegraphics[width=0.7\linewidth,angle=0]{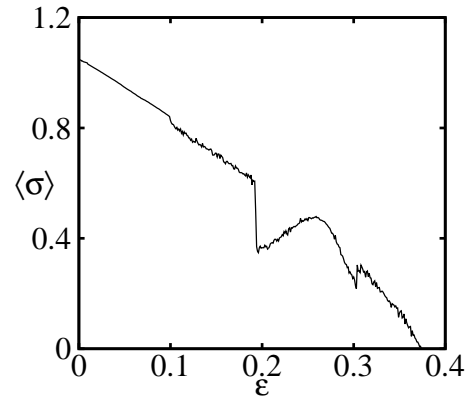}
\end{center}
\caption{$\langle \sigma \rangle$ as a function of the coupling strength $\epsilon$ for the coupled robust-chaos system Eq.~(\ref{Homo}), with
size $N=10^4$ and fixed local parameters  $a=0.375$, $b=0.826$.
A transient time of $3 \times 10^3$ integration steps were discarded and the subsequent $7 \times 10^3$ steps were used
to calculate $\langle \sigma \rangle$ for each value of $\epsilon$.}
\label{F2}     
\end{figure}

Figure~(\ref{F2}) shows the quantity $\langle \sigma \rangle$ computed as a function of the coupling parameter $\epsilon$. We observe that the 
system Eq.~(\ref{Homo}) becomes synchronized at the value  $\epsilon=0.375=a$.

\begin{figure}[h]
\begin{center}
\includegraphics[width=0.7\linewidth,angle=0]{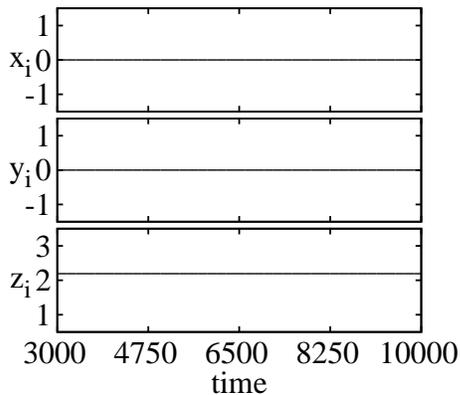}
\end{center}
\caption{Variables $x_i$, $y_i$, $z_i$ for one oscillator in the system Eq.~(\ref{Homo}) with $\epsilon=a=0.375$ as  functions of $t$, 
after $3000$ transients.  The synchronization value $z_i=z^*=k$ depends on initial conditions; here $k=2.19$.
}
\label{F3}
\end{figure}

To elucidate the nature of the synchronized state observed in Fig.~(\ref{F2}), we have plotted in Fig.~(\ref{F3}) the asymptotic values of the 
variables $x_i$, $y_i$, and $z_i$ for one oscillator in the system Eq.~(\ref{Homo}) with $\epsilon=a=0.375$, as functions of time.
The dynamics of the oscillator has reached the fixed point $x^*=0,  y^*=0,  z^*=2.19$, which is evidently stable. 
Since for this value of $\epsilon$ the system is synchronized, all oscillators share this fixed point state. Thus, we have evidence
for the occurrence of amplitude death in the coupled robust-chaos system Eq.~(\ref{Homo}). 
The global coupling induces suppression of the chaotic dynamics in the system.

In many real systems, the interacting dynamical elements are not identical. Thus, in general it is important to investigate the effects of
heterogeneities in the local dynamics. In this respect, the existence of an extensive region of parameters exhibiting robust chaos 
in the local dynamics, Eq.~(\ref{Shi}), allows to explore the behavior of coupled heterogeneous, robust-chaos flows. Then, we consider
the system of heterogeneous oscillators,
\begin{eqnarray}
\label{Hetero}
 \frac{dx_i}{dt} &=& y_i  ,\nonumber \\
  \frac{dy_i}{dt} &=& x_i(1-z_i)-b_iy_i  , \\
   \frac{dz_i}{dt} &=& -a_i(z_i-x_i^2)  +\epsilon \bar{z}  , \nonumber 
\end{eqnarray}
where the parameter values $a_i$ are assigned from a random and uniform distribution on the robust chaos interval $a_i \in [0.1, 0.4]$,
and the corresponding parameters $b_i$ are fixed by Eq.~(\ref{recta}), i. e., $b_i=1.67a_i+0.2$. 

\begin{figure}[h]
\begin{center}
\includegraphics[width=0.325\linewidth,angle=0]{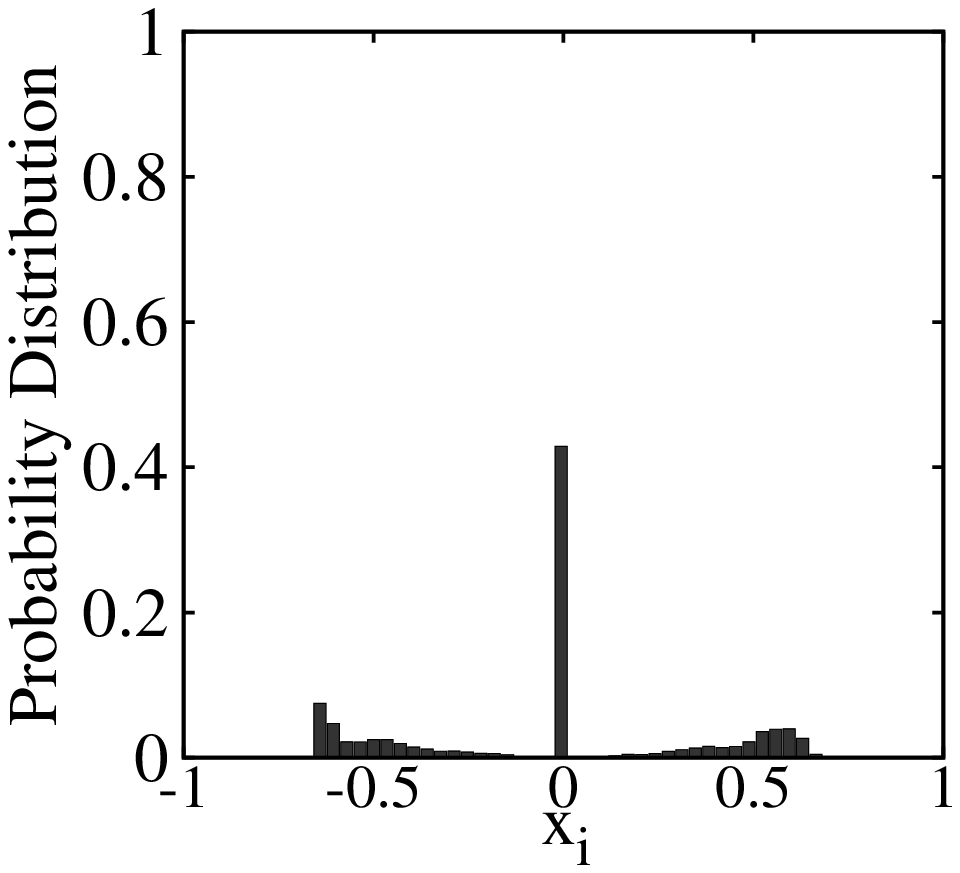}
\includegraphics[width=0.325\linewidth,angle=0]{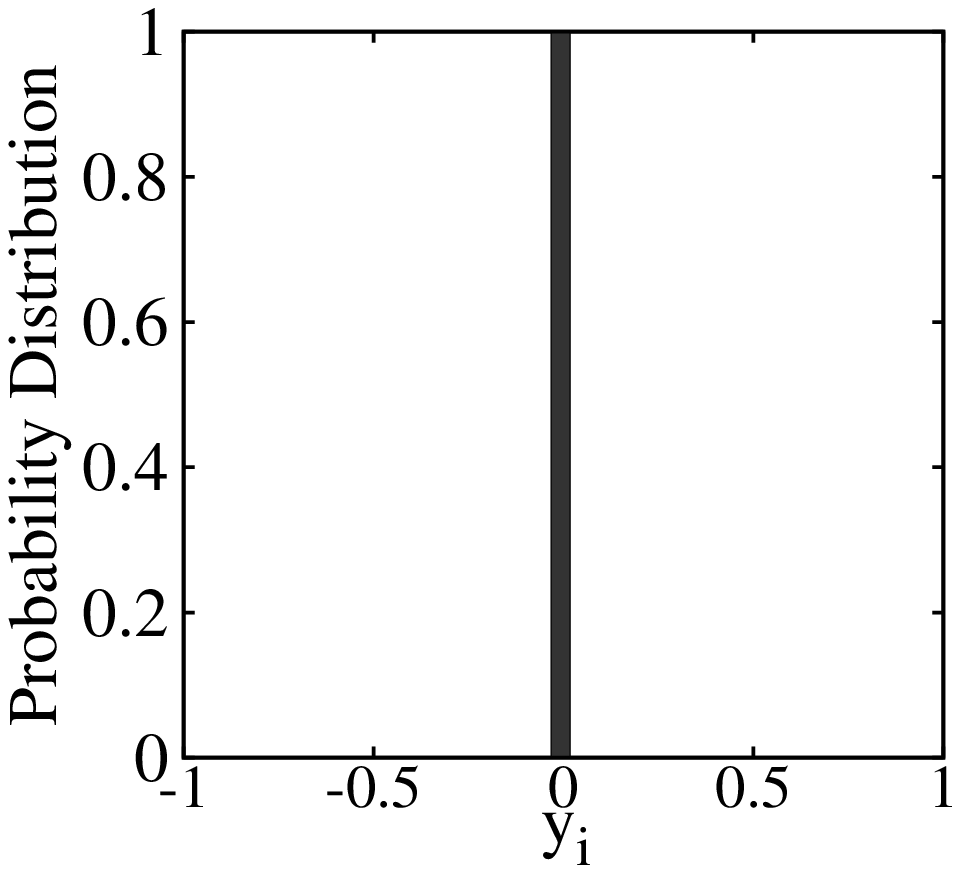}
\includegraphics[width=0.325\linewidth,angle=0]{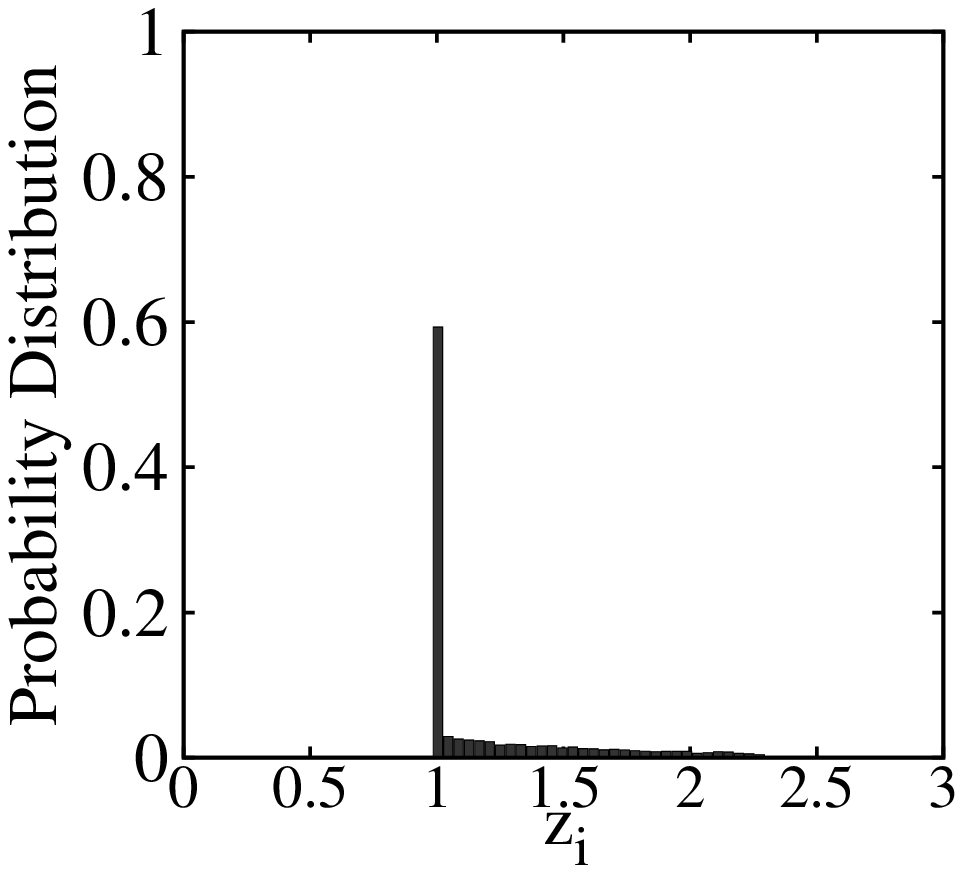}
\end{center}
\caption{Probability distributions of the stationary state variables for the heterogeneous robust-chaos system Eq.~(\ref{Hetero}), with $N=10^5$,
$\epsilon=0.2$ at a time $t=10^4$. 
}
\label{F4}
\end{figure}

The system Eq.~(\ref{Hetero}) does not synchronize for any value of the coupling parameter $\epsilon$ 
(for large enough $\epsilon$, the trajectories of this system diverge). Instead, we observe the
convergence of the orbits the oscillators to different stationary states. Figure~(\ref{F4}) shows the probability distributions of state variables
$x_i$, $y_i$, and $z_i$ for the oscillators in system Eq.~(\ref{Hetero}) at an asymptotic time. We see that $y_i=0$, $\forall i$; however
the stationary values of the variables $x_i$ and $y_i$ are distributed over a respective interval; giving raise to the formation of a large cluster in the fixed point  $(x^*=0, y^*=0, z^*=1)$, and several other small stationary clusters.

In this case, the cessation of the oscillations gives place to several coexisting steady states, and a collective inhomogeneous steady 
state appears. This phenomenon is referred to as oscillation death, in contrast to situation occurring in  amplitude death, where 
the suppression of oscillations in the system is manifested through the presence of a homogeneous steady state \cite{Koseska}.

In summary, we have investigated the synchronization behavior of a system of globally coupled robust-chaos, continuous-time oscillators. 
The local dynamics corresponds to the Shimizu-Morioka model where the occurrence of robust chaos in a region of its 
parameter space was recently found \cite{Gallas}.  

We have showed that the global coupling can drive the oscillators to synchronization into a fixed point created by the coupling, 
resulting in amplitude death. 
We have also investigated the influence of heterogeneity, which is a common feature in many real systems of interacting elements. 
The existence of robust chaos allows to introduce heterogeneity in the local parameters,
while guaranteeing the functioning of all the oscillators in a chaotic mode. 
In this case, we found that the system reaches
a state of oscillation death, with coexisting clusters of oscillators in different steady states. 
The phenomena of amplitude death or oscillation death in coupled robust-chaos flows could be employed as mechanisms 
for stabilization and control routes, as for example, in a realistic array of reliable chaotic lasers dedicated to secure communications.

Amplitude death has been considered as a general phenomenon realized for periodic, chaotic, hyperchaotic, time delayed systems, 
and also for various coupling schemes, i.e. diffusive, replacement, conjugate coupling, coupling via scalar signals, nonlinear 
coupling, etc. \cite{Rama2,Koseska}. Our results permit to add coupled robust-chaos oscillators to the list of systems exhibiting both
amplitude and oscillation death.  

Future research work on systems of coupled robust-chaos flows should include the investigation of nontrivial collective behavior, i.e., 
the periodic evolution of macroscopic quantities coexisting with local chaotic dynamics, a phenomenon found in networks of
coupled robust-chaos maps \cite{Javier,Alvarez}.

\subsection*{Acknowledgments}
This work is supported by project No. C-1827-13-05-B from CDCHTA, 
Universidad de Los Andes, Venezuela. M. G. C. is grateful to the Senior Associates Program of 
the Abdus Salam International Centre for Theoretical Physics, Trieste, Italy, for the visiting opportunities.

\end{document}